\documentclass{jps-cp}
\usepackage{times}

\title{Two-Channel Kondo Effect Emerging from Np and Pu Ions}

\author{Dai \textsc{Matsui} and Takashi \textsc{Hotta}}

\inst{Department of Physics, Tokyo Metropolitan University, Tokyo 192-0397, Japan}

\email{hotta@tmu.ac.jp}

\recdate{September 16, 2019}

\abst{
On the basis of an idea of the electron-hole relation between $f^2$ and $f^4$
states in a $j$-$j$ coupling scheme,
we point out a possibility that two-channel Kondo phenomena
can be observed even in cubic transuranium compounds
including Np and Pu ions
with local $\Gamma_3$ non-Kramers doublet ground state.
For the purpose, we analyze a seven-orbital impurity Anderson model
hybridized with degenerate $\Gamma_8$ conduction electron bands
by employing a numerical renormalization group method.
From the numerical results for the case of four local $f$ electrons,
corresponding to Np$^{3+}$ and Pu$^{4+}$ ions,
we confirm that a residual entropy of $0.5\log 2$,
a characteristic of two-channel Kondo phenomena,
appears for the case with the local $\Gamma_3$ doublet ground state.
}

\kword{Two-channel Kondo effect,  transuranium compounds, $j$-$j$ coupling scheme, numerical renormalization group method}

\begin{document}
\maketitle

\section{Introduction}

It has been well known that the Kondo effect occurs in a dilute magnetic impurity system
\cite{Kondo1,Kondo2a,Kondo2b}.
For the case in which a single impurity spin $1/2$ is embedded in
a single-band conduction electron system, we observe conventional Kondo phenomena
and the mechanism has been completely understood both from theoretical and
experimental viewpoints \cite{Kondo40}.
On the other hand, for the case in which plural numbers of electrons
with spin-orbital complex degrees of freedom are hybridized with
multi-channel conduction electron bands,
new and rich phenomena have been actively discussed for a long time,
after the clarification of the conventional Kondo effect.

When an impurity spin is hybridized with multi-channel conduction bands,
an intriguing concept of multi-channel Kondo effect has been proposed \cite{Nozieres}.
In particular, for the case of impurity spin $1/2$ and two conduction bands,
the appearance of a non-Fermi liquid ground state has been suggested.
This is one of the fascinating properties of two-channel Kondo effect.
Such a non-Fermi liquid state has been also pointed out
in a two-impurity Kondo system \cite{Jones1,Jones2}.
Concerning the reality of the two-channel Kondo effect,
Cox has pointed out that two screening channels exist
in the case of quadrupole degree of freedom in a cubic uranium compound
with $\Gamma_3$ non-Kramers doublet ground state \cite{Cox1,Cox2}.
After this proposal, the stage for the investigation of two-channel Kondo
phenomena has been almost fixed as the $f^2$ electron system.
First, cubic U compounds have been focused, but recently,
cubic Pr compounds have been actively investigated \cite{review}.

Although it is important to investigate the $f^2$ electron system
to deepen our understanding on the two-channel Kondo effect,
we strongly believe that it is also meaningful to expand the research frontier
of the two-channel Kondo physics to other rare-earth and actinide compounds.
On the basis of our belief, we have examined the Kondo effect for the case of
plural numbers of $f$ electrons in rare-earth and actinide compounds
by analyzing a seven-orbital impurity Anderson model hybridized with
$\Gamma_8$ conduction bands with the use of a numerical renormalization
group (NRG) method \cite{NRG}.
First we have reconfirmed the appearance of quadrupole two-channel Kondo
effect for the case of $n=2$ \cite{Hotta1},
where $n$ denotes the local $f$ electron number per ion.
Then, we have moved onto the case of $n=3$,
corresponding to Nd ion,
in which we have clarified the emergence of magnetic two-channel Kondo
effect for the case with a $\Gamma_6$ ground state \cite{Hotta2}.

In this paper, we analyze the seven-orbital impurity Anderson model
hybridized with $\Gamma_8$ conduction electrons by using the NRG method
for the case of $n=4$, corresponding to Np$^{3+}$ and Pu$^{4+}$ ions. 
We confirm that a residual entropy of $0.5 \log 2$ appears
as a clear signal of  the quadrupole two-channel Kondo effect
for the case with the non-Kramers $\Gamma_3$ doublet ground state.
In addition, we also find a quantum critical point (QCP) between
local crystalline electric field (CEF) singlet and
Kondo-Yosida singlet states for the case of $n=4$,
as has been found in the case of $n=2$.
Finally, we briefly discuss potential materials to observe actually
the quadrupole two-channel Kondo effect for the case of $n=4$.
Throughout this paper, we use such units as $\hbar=k_{\rm B}=1$.

\section{Model and Method}

To describe the local $f$-electron model,
first we consider one $f$-electron state,
which is the eigenstate of spin-orbit and CEF terms.
Under the cubic CEF potential,
$\Gamma_7$ doublet and $\Gamma_8$ quartet are obtained
from $j=5/2$ sextet,
whereas we obtain $\Gamma_6$ doublet, $\Gamma_7$ doublet,
and $\Gamma_8$ quartet from $j=7/2$ octet,
where $j$ denotes the total angular momentum of $f$ electron.
By using those one-electron states as bases,
we express the local $f$-eletctron Hamiltonian as
\begin{equation}
\label{Hloc}
\begin{split}
  H_{\rm loc} &= \sum_{j, \mu, \tau} (E_f+\lambda_j  + B_{j,\mu})
  f_{j \mu \tau}^{\dag} f_{j \mu \tau} \\
  &+\sum_{j_1\sim  j_4} \sum_{\mu_1 \sim \mu_4} 
  \sum_{\tau_1 \sim \tau_4} 
  I^{j_1 j_2, j_3 j_4}_{\mu_1 \tau_1 \mu_2 \tau_2, \mu_3 \tau_3 \mu_4 \tau_4}
  f_{j_1 \mu_1 \tau_1}^{\dag} f_{j_2 \mu_2 \tau_2}^{\dag}
  f_{j_3 \mu_3 \tau_3}  f_{j_4 \mu_4 \tau_4},
\end{split}
\end{equation}
where $E_f$ is the $f$-electron level to control the local $f$-electron number $n$ at an impurity site,
$f_{j \mu\tau}$ denotes the annihilation operator of a localized $f$ electron in the bases of
$(j, \mu, \tau)$, $j=5/2$ and $7/2$ are denoted by ``$a$'' and ``$b$'', respectively,
$\mu$ distinguishes the cubic irreducible representations,
$\Gamma_8$ states are distinguished by $\mu=\alpha$ and $\beta$,
while $\Gamma_7$ and $\Gamma_6$ states are labeled by $\mu=\gamma$ and $\delta$, respectively,
and $\tau$ is the pseudo-spin which distinguishes the degeneracy concerning the time-reversal symmetry.
The definitions of $\lambda_j$, $B_{j,\mu}$, and $I$ will be discussed below.

As for the spin-orbit term, we obtain
\begin{equation}
 \lambda_a=-2\lambda,~\lambda_b=(3/2)\lambda,
\end{equation}
where $\lambda$ is the spin-orbit coupling of $f$ electron.
The magnitude of $\lambda$ depends on the kind of actinide atoms,
but in this paper, we set $\lambda=0.3$ eV,
since we consider Np and Pu ions.

Concerning the CEF potential term for $j=5/2$,
we obtain
\begin{equation}
B_{a,\alpha} =B_{a,\beta}=1320 B_4^0/7,~B_{a,\gamma} =-2640 B_4^0/7,
\end{equation}
where $B_4^0$ denotes the fourth-order CEF parameter
in the table of Hutchings for the angular momentum $\ell=3$
\cite{Hutchings,LLW}.
Here we note that the sixth-order CEF potential term $B_6^0$
does not appear for $j=5/2$, since the maximum size
of the change of the total angular momentum is less than six.
On the other hand, for $j=7/2$, we obtain
\begin{equation}
\begin{split}
B_{b,\alpha} &=B_{b,\beta}=360B_4^0/7+2880B_6^0,\\
B_{b,\gamma} &=-3240B_4^0/7-2160 B_6^0,\\
B_{b,\delta} &=360 B_4^0-3600 B_6^0/7.
\end{split}
\end{equation}
Note also that $B_6^0$ terms appear in this case.
In the present calculations, we treat $B_4^0$ and $B_6^0$
as parameters.

The matrix element of the Coulomb interaction is expressed by $I$.
To save space, we do not show the explicit forms of $I$ here,
but they are expressed by four Slater-Condon parameters
($F^0$, $F^2$, $F^4$, $F^6$) \cite{Slater,Condon}
and Gaunt coefficients \cite{Gaunt,Racah}.
As for the magnitudes of Slater-Condon parameters,
first we set $F^0=10$ eV by hand.
Others are determined so as to reproduce excitation spectra of
U$^{4+}$ ion with two $5f$ electrons \cite{Eliav}.
The results are 
$F^2=6.4$ eV, $F^4=5.6$ eV, and $F^6=4.1$ eV \cite{Hotta3}.

Now we include the $\Gamma_8$ conduction electron bands
hybridized with localized electrons.
Since we consider the case of $n <7 $, the local $f$-electron states are mainly
formed by $j=5/2$ electrons and the chemical potential is situated
among the $j=5/2$ sextet.
Thus, we consider only the hybridization between conduction and
$j=5/2$ electrons.
The seven-orbital impurity Anderson model is given by
\begin{equation}
  H = \sum_{\mib{k},\mu,\tau} \varepsilon_{\mib{k}}
   c_{\mib{k}\mu\tau}^{\dag} c_{\mib{k}\mu\tau}
   + \sum_{\mib{k},\mu,\tau} V_{\mu}(c_{\mib{k}\mu\tau}^{\dag}f_{a\mu\tau}+{\rm h.c.})
   +  H_{\rm loc},
\end{equation}
where $\varepsilon_{\mib{k}}$ is the dispersion of conduction electron with wave vector $\mib{k}$,
$c_{\mib{k}\gamma\tau}$ denotes an annihilation operator of conduction electron,
and we set $V_{\alpha}=V_{\beta}=V$.
Here $V$ denotes the hybridization between $\Gamma_8$ conduction and localized electrons.

In this paper, we analyze the model by employing the NRG method \cite{NRG}.
We introduce a cut-off $\Lambda$ for the logarithmic discretization of
the conduction band.
Due to the limitation of computer resources,
we keep $M$ low-energy states.
Here we use $\Lambda=5$ and $M=2,500$.
Note that the temperature $T$ is defined as
$T=\Lambda^{-(N-1)/2}$ in the NRG calculation,
where $N$ is the number of the renormalization step.
In the following calculation, the energy unit is $D$, which is
a half of conduction band width.
Namely, we assume $D=1$ eV in this calculation.

\begin{figure}[t]
\includegraphics[width=0.75\linewidth]{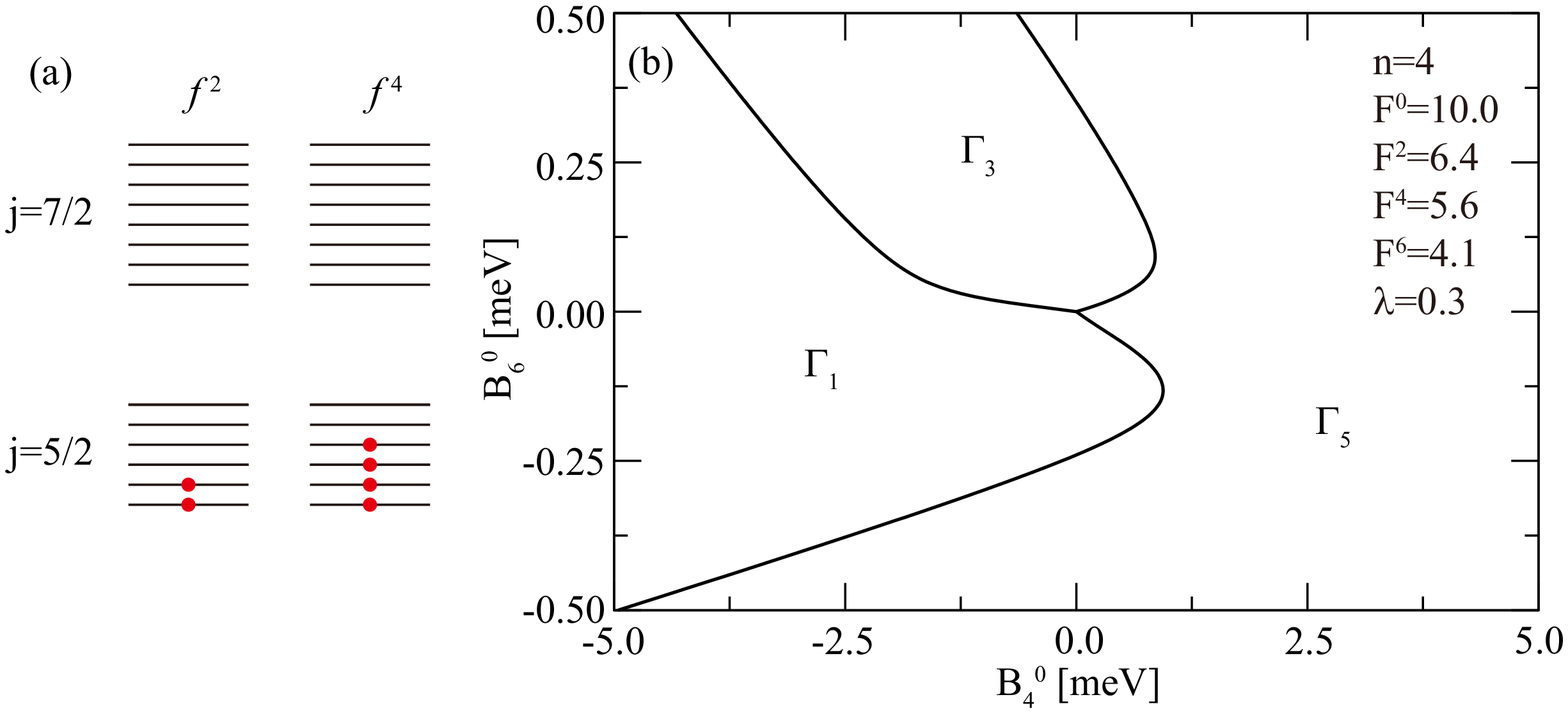}
\caption{
(a) Electron configurations in the $j$-$j$ coupling scheme for $n=2$ and $n=4$.
(b) Ground-state phase diagram of $H_{\rm loc}$
on the $(B_4^0, B_6^0)$ plane for $n=4$.
}
\end{figure}

\section{Calculation Results}

First we briefly discuss the $f$-electron configurations in a $j$-$j$ coupling
scheme for $n=2$ and $n=4$.
As shown in Fig.~1(a), when we accommodate four $f$ electrons
in the $j=5/2$ sextet, we find two $f$ holes there.
Of course, for the finite value of $\lambda$, the component of
$j=7/2$ octet should be included in the ground-state wave function.
Note here that we mention only the main component of the
ground and low-energy excited states for $n=2$ and $n=4$
\cite{Hotta4,Hotta5}.
We emphasize that the electron-hole relation between $n=2$ and $n=4$
on the basis of the $j$-$j$ coupling scheme is quite simple,
but we expect to observe the quadrupole two-channel Kondo effect
even for $n=4$, corresponding to Np$^{3+}$ and Pu$^{4+}$ ions.

Now we consider the local CEF ground-state phase diagram
for $n=4$, obtained from the diagonalization of $H_{\rm loc}$.
For $n=4$,  the ground-state multiplet for $B_4^0=B_6^0=0$ is
characterized by total angular momentum $J=4$.
Under the cubic CEF potentials, the nonet of $J=4$ is split into four
groups as $\Gamma_1$ singlet, $\Gamma_3$ doublet, $\Gamma_4$ triplet,
and $\Gamma_5$ triplet.
Among them, $\Gamma_4$ triplet does not appear as a solo
ground state under the cubic CEF potential
with $O_{\rm h}$ symmetry.

Then, we obtain three local ground states for $n=4$, as shown in Fig.~1(b).
Roughly speaking, we obtain $\Gamma_1$ singlet for $B_4^0<0$,
whereas $\Gamma_5$ triplet appears for $B_4^0>0$.
Here we recall the fact that $f^1$ local ground state
is $\Gamma_7$ and $\Gamma_8$ for $B_4^0>0$ and $B_4^0<0$,
respectively, from eq.~(3).
When we accommodate two holes into this situation,
we easily obtain $\Gamma_1$ singlet and $\Gamma_5$ triplet
for $B_4^0<0$ and $B_4^0>0$, respectively,
by standard positive Hund's rule coupling.
Note that the results are just reversed in comparison with the
the case of $n=2$, in which we have found $\Gamma_1$ singlet
and $\Gamma_5$ triplet for $B_4^0>0$ and $B_4^0<0$,
respectively \cite{Hotta1},
since the signs in the one-electron potentials are changed between
the electron and hole pictures.
As for $\Gamma_3$ doublet, it appears for $B_6^0>0$ near
the region of $B_4^0 \approx 0$.
The stabilization of $\Gamma_3$ doublet is understood
by effective negative Hund's rule coupling
and the dependence on $B_6^0$ for the hole picture
is the same as that for the electron one.

\begin{figure}[t]
\includegraphics[width=0.67\linewidth]{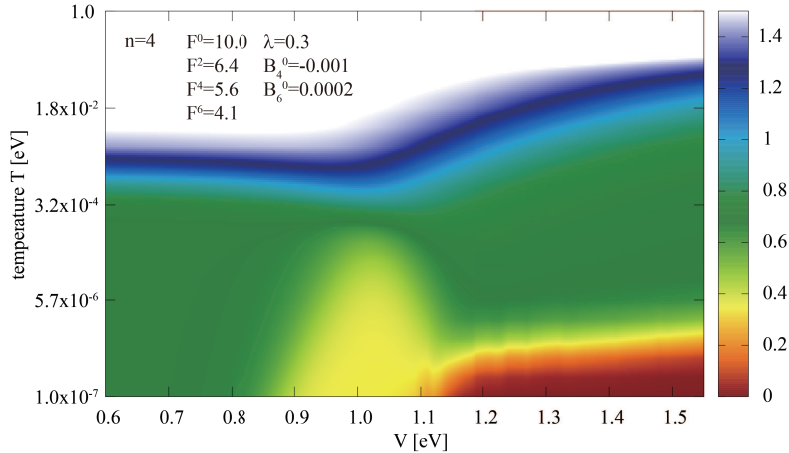}
\caption{Color contour map of entropy for $n=4$
on the $(V, T)$ plane for $B_4^0=-0.001$ and $B_6^0=0.0002$.
Note that $T$ is given in a logarithmic scale.
}
\end{figure}

Next we show our NRG results.
First we discuss the $V$ dependence of the
two-channel Kondo phase for the present parameters.
In Fig.~2, we show the results for entropy on the $(V,T)$ plane
for $B_4^0=-0.001$ and $B_6^0=0.0002$ with the $\Gamma_3$ local ground state.
To visualize precisely the behavior of entropy,
we define the color of the entropy between $0$ and $1.5$,
as shown in the right color bar.
Note that the green and yellow regions indicate
the entropy of $\log 2$ and $0.5 \log 2$, respectively.
We emphasize  that $0.5\log 2$ entropy does not appear
only at a certain value of $V$,
but it can be observed in the wide region of $V$
as $0.9 < V < 1.1$ in the present temperature range.
This behavior is different from that in the non-Fermi liquid state
due to the competition between CEF and Kondo-Yosida singlets,
as will be discussed later.
Note also that the two-channel Kondo effect appears for relatively
large values of $V$ in the present energy scale of $D=1$ eV.
In the following calculations, we set $V=1$ eV.

\begin{figure}[t]
\includegraphics[width=0.67\linewidth]{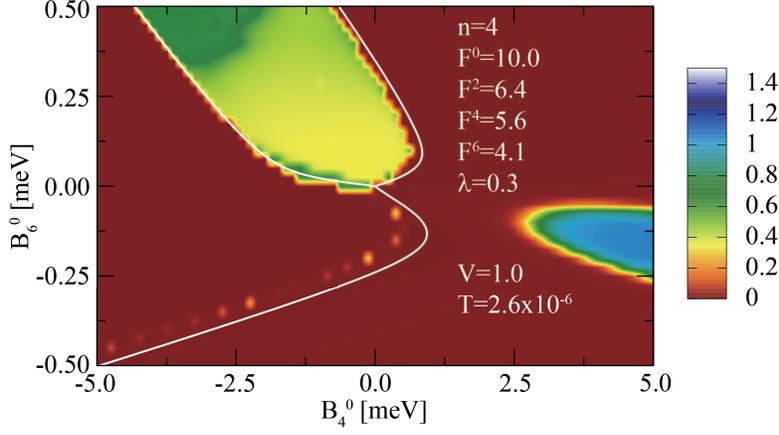}
\caption{Color contour map of the entropy for $n=4$ on
the plane of ($B_4^0$, $B_6^0)$ for $V$=1.0 and $T=2.6 \times 10^{-6}$.
White curves denote the boundaries among local CEF ground states
shown in Fig.~1.
Note that yellow spots appear along the boundary curve between
$\Gamma_1$ and $\Gamma_5$ regions.}
\end{figure}

In Fig.~3, we show the contour color map of the entropy
on the plane of ($B_4^0$, $B_6^0)$
for $V=1.0$ and $T=2.6 \times 10^{-6}$.
The definition of the color is the same as that in Fig.~2.
We immediately notice that a region with an entropy of $0.5\log 2$
(yellow region) almost corresponds to that of the $\Gamma_3$
ground state in comparison with Fig.~1(b).
Note that for large $B_6^0$, the color becomes green,
but when we decrease the temperature, we find the entropy of  $0.5\log 2$
even for large $B_6^0$.
The results strongly suggest the emergence of quadrupole two-channel
Kondo effect for the case of $n=4$.
Note that in the present model, we have observed the discontinuous
change in entropy behavior between yellow ($\Gamma_3$) and red
($\Gamma_1$ or $\Gamma_5$) regions.
Quantum critical behavior which may occur between non-Fermi-liquid and
Fermi-liquid phases will be discussed elsewhere in future.

Here we remark that there exists a blue region corresponding to $\log 3$
in the local $\Gamma_5$ state.
This is quite natural, since the local $\Gamma_5$ state is
triply degenerate.
The $\Gamma_5$ moment is screened by $\Gamma_8$ conduction electrons
and thus, it is expected that the conventional Kondo effect occurs in
the $\Gamma_5$ region,
although the magnitude of the Kondo temperature significantly
depends on the hybridization and excitation energy.

\begin{figure}[t]
\includegraphics[width=0.55\linewidth]{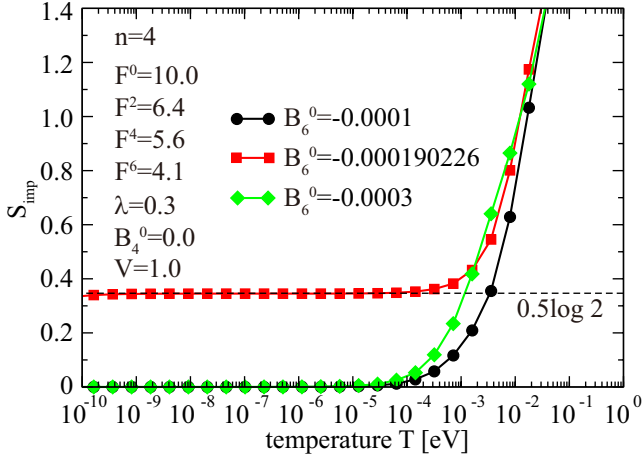}
\caption{
Entropies on the line of $B_4^0=0$
for $B_6^0=-0.0001$, $-0.000190226$, and $-0.0003$.
}
\end{figure}

Note also that in the region of $B_6^0 <0$, we observe some
blurry yellow spots along the boundary curve between
$\Gamma_1$ singlet and $\Gamma_5$ triplet local ground states.
The QCP is known to appear
between the local CEF and Kondo singlet states for $n=2$ \cite{Hotta1}.
Thus, we deduce that those spots form a QCP curve, although we could not
obtain enough amounts of numerical results to depict the smooth curve.

In Fig.~4, we show the curves of entropy vs. temperature
for the points near $B_4^0=0.0$ and $B_6^0=-0.0002$.
For $B_6^0=-0.0001$ (CEF singlet) and $-0.0003$ (Kondo singlet),
entropies are found to be zero around at $T=10^{-4}$,
while we find a residual entropy of $0.5 \log 2$ even at $T=10^{-10}$
for $B_6^0=-0.000190226$.
When $B_6^0$ deviates even slightly from this value,
we find that $0.5 \log 2$ entropy immediately disappears.
The present results are quite similar to those for $n=2$,
which we have actually found in the same model \cite {Hotta1}.
In this sense, we highly expect the existence of QCP
between CEF and Kondo singlet states even in the $f^4$ system.


\section{Summary and Discussion}

In this paper, we have analyzed the seven-orbital impurity Anderson model
hybridized with $\Gamma_8$ conduction bands by using the NRG technique.
For the case of $n=4$, we have confirmed the emergence of
two-channel Kondo effect.
As for the mechanism of quadrupole two-channel
Kondo effect for the case of $n=4$,
we deduce that it is essentially the same as that
for the case of $n=2$ from the viewpoint of
the electron-hole relation between $f^2$ and $f^4$ states
on the basis of the $j$-$j$ coupling.

We have also observed the QCP curve running in the $\Gamma_1$ region
near the boundary between the CEF singlet and Kondo-Yosida singlet states.
Here readers may have some questions on this QCP curve.
For instance, the curve seems to merge into the two-channel Kondo state in
the local $\Gamma_3$ region,
but the property of the state at the merging point is unclear.
The details on the properties of the QCP curve will be discussed
elsewhere in future.

Finally, we provide a brief comment on actual materials to observe
the two-channel Kondo effect for the case of $n=4$.
In rare-earth ions, the case of $n=4$ corresponds to Pm$^{3+}$,
but unfortunately, there exist no stable isotopes for Pm.
Thus, we turn our attention to actinide ions such as Np$^{3+}$ and Pu$^{4+}$
with $5f^4$ configurations. It may be difficult to synthesize
new Np and Pu compounds, but we expect that
Np 1-2-20 compound will be synthesized in future.

\section*{Acknowledgement}

This work was supported by JSPS KAKENHI Grant Number JP16H04017. 
The computation in this work was done using the facilities of the
Supercomputer Center of Institute for Solid State Physics, University of Tokyo.


\end{document}